\documentclass[twocolumn,runningheads]{svjour2}
\smartqed  
\usepackage{graphicx}
\usepackage{mathptmx}      

\newcommand{\psrcrab}{PSR~J0534+2200}

\newcommand{\psrmsh}{PSR~J1513-5908}

\newcommand{\psrvela}{PSR~J0835-4510}
\newcommand{\vela}{Vela~pulsar}
\newcommand{\psrkooka}{PSR~J1420-6048}
\newcommand{\kooka}{Kookaburra}
\newcommand{\inte}{\textsl{INTEGRAL}}
\newcommand{\rosat}{\textsl{ROSAT}}
\newcommand{\hess}{\textsl{H.E.S.S.}}
\newcommand{\ibis}{\textsl{IBIS (ISGRI)}}
\journalname{Astrophysics and Space Science}
\begin{document}

\title{INTEGRAL observations of TeV plerions
}

\author{A.I.D. Hoffmann         \and
        D. Horns \and
        A. Santangelo
}

\institute{A.I.D. Hoffmann \and D. Horns \and A. Santangelo \at
  Institut f\"ur Astronomie und Astrophysik, Universit\"at T\"ubingen, Germany \\
  Tel.: +49 7071 29 76132 \\
  Fax:  +49 7071 29 3458 \\
              \email{hoffmann@astro.uni-tuebingen.de}           
}

\date{Received: date / Accepted: date}

\maketitle

\begin{abstract}
  Amongst the sources seen in very high gamma-rays several are associated
  with Pulsar Wind Nebulae (``TeV plerions''). The study of hard X-ray/soft
  gamma-ray emission is providing an 
  important insight into the energetic particle population present in these
  objects. The unpulsed emission from pulsar/pulsar wind nebula systems in
  the energy range accessible to the \inte\ satellite is mainly
  synchrotron emission from energetic and fast cooling electrons close to
  their acceleration site.  Our analyses of public \inte \ data of known
  TeV plerions detected by ground based Cherenkov telescopes indicate a
  deeper link between these TeV plerions and \inte \ detected pulsar wind
  nebulae. The newly discovered TeV plerion in the northern wing of the
  Kookaburra region (G313.3+0.6 powered by the middle aged \psrkooka) is found
  to have a previously unknown \inte \ counterpart which is besides the
  \vela \ the only middle aged pulsar detected with \inte. We do not
  find an \inte \ counterpart of the TeV plerion associated with the X-ray
  PWN ``Rabbit'' G313.3+0.1 which is possibly powered by a young pulsar.
\end{abstract}
\section{Introduction - What are Rotation powered pulsar wind nebulae (RPWN)?}
\label{intro}
Rotation powered pulsar wind nebulae (RPWN) contain an isolated neutron star
that drives a relativistic wind of particles into the ambient medium fueling
an extended non-thermal emission region.  \\
The rotational energy of the neutron star is partially transformed into a
highly relativistic wind of particles that is driven~ into the ambient
medium which~ (for middle aged RPWN) is potentially modified by the reverse
shock of the supernova remnant ejecta \cite{Blondin01}.  A relativistic
standing shock forms at a distance of typically $10^{16}-10^{17}$~cm to the
central object and is believed to be the site of particle acceleration.
Among the possible ways of accelerating particles, the well-known Fermi
acceleration mechanism has been considered to be a good candidate to
explain the broad band power law distribution of the pair plasma in the
downstream region. 

\section{Particle acceleration in RPWN}
\label{sec:1}
The ultrarelativistic wind with bulk Lorentz factors of about $\gamma =
10^4 - 10^7$ terminates in a standing shock. The details of Fermi-type
shock acceleration in relativistic shocks has become recently a matter of
controversy. While initial calculations
\cite{KirkSchneider87,Achterberg01,Ostrowski91} indicated an efficient
acceleration of particles following a universal power law type
distribution, recent calculations have revealed substantial difficulties in
accelerating particles to form a power law type particle distribution
\cite{NiemiecOstrowski06,Lemoine06}. Moreover, detailed particle in cell
distributions of shocks forming in flows with different magnetization have
not been found to show any acceleration at all \cite{Spitkovsky06}. While
the issue is currently not settled, it appears worthwhile to consider
alternative mechanisms to explain the acceleration of particles in
ultrarelativistic shocks. An interesting alternative has been proposed by
Hoshino et al. \cite{Hoshino92}, and Arons \& Tavani \cite{AronsTavani94}
in which in the downstream region gyrating, reflecting ions dissipate
energy in magnetosonic waves which are absorbed by the pair-plasma to form
a power law type spectrum. An important prediction of this model is that
the maximum energy achievable in this type of acceleration is given by
$\gamma m_i c^2 /Z $, where $ m_i$ indicates the mass and $Z$ the charge of
the ions.
\section{Hard X-ray emission from RPWN: \\ In situ tracer of particle acceleration}
\label{sec:2}
While the archetypal RPWN, the Crab nebula is a well studied example of a
young ($t\ll 10$~kyrs) RPWN which is sufficiently bright to measure the
emitted power over an extremely wide energy range, similar detailed energy
spectra for middle aged RPWN ($t\approx 10$~kyrs) have not been obtained.
The hard X-ray band ($20-100$~keV) and the gamma-ray band ($0.1-100$~MeV)
are crucial observational windows to explore the acceleration of particles:
In this energy range, we expect predominantly synchrotron emission of
accelerated electrons. Moreover, the life-time of the electrons emitting in
the hard X-ray band is rather limited
\begin{equation}
  t_{1/2}\approx 8.4~\mathrm{yrs} (B_{-4})^{-3/2} (\epsilon/20~\mathrm{keV})^{-1/2}
\end{equation}
with $B=B_{-4}\cdot 10^{-4}\mathrm{G}$. In general, we expect unpulsed hard
X-rays to be emitted only in a very confined volume close to the
aceleration site.
\section{INTEGRAL observations of TeV plerions}
\label{sec:3}
The \inte \ instruments have been used to observe regularly a large
fraction of the Galactic disk. A rather small fraction  of a few per cent
of the sources detected by the \inte \ satellite have been identified to be 
young and middle-aged RPWN
(see Tab. \ref{tab:pulsarparameter}). We have analysed part of the
archival data focussing on RPWN which have been detected as TeV plerions.
In addition to already known \inte \ sources, we find evidence for hard
X-ray emission from the newly discovered ``Kookaburra'' TeV plerion G313.3+0.6.

\begin{figure}[t]
  \centering
  \includegraphics[width=0.95\linewidth]{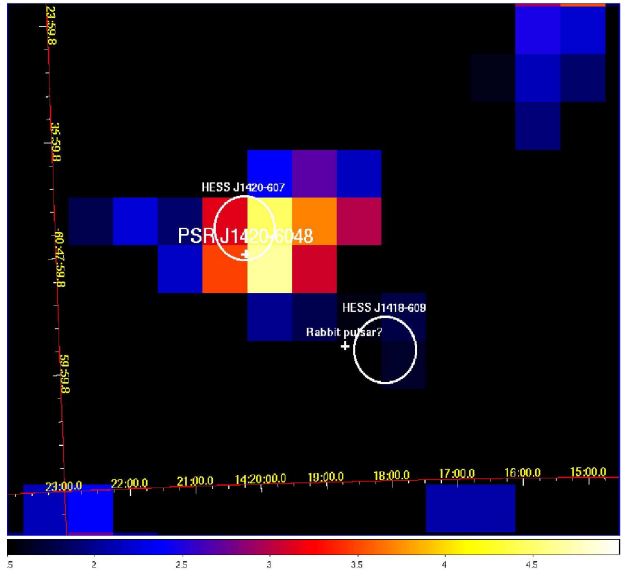}
  \caption{  \label{fig:kookainte} $20-40$ keV significance map (\ibis \
    data, $\approx$ 940 ksec exposure) of the Kookaburra region. The gray
    scale is truncated below $+1.5\sigma$. The analysis was done with OSA
    5.0 for all public \inte \ data up to a maximal distance to the source
    position of $10.0^{\circ}$.}
\end{figure}

Here we present results from two mosaic images assembled of public \inte \ 
observations. Both mosaics were generated using the standard \inte \ 
offline analysis package OSA 5.0. The mosaic centered on the region of the
``Kookaburra'' incorporates all public \inte \ data up to a maximal
distance of $10^{\circ}$ to the source position of the Kookaburra region.
While Fig. \ref{fig:kookainte} shows this region, Fig. \ref{fig:mshinte} is
a section of the same mosaic showing \psrmsh, which due to the large FOV of
\inte \ is also in this mosaic. But there are more public data for this
pulsar available in the archive. The total exposure time for this mosaic is
about $940$~ksec.\\
The other mosaic (Fig. \ref{fig:velainte}) uses all public \inte \ data
available for a maximal distance of $4.5^{\circ}$ from the position of the
\vela.  The total exposure time for this mosaic is $\approx~1.3$~Msec.

\begin{table*}[t!]
  \centering
  \caption{Some parameters of Pulsars seen with \inte \ and \hess. \psrcrab
    \ and \psrmsh \ are young pulsars, \psrvela \ and \psrkooka \ are
    middle-aged pulsars. All values except VHE and
    X-ray luminosities from Manchester et al. (2005) (\cite{Manchester05}.) 
    The VHE luminosities are calculated based on the power law 
    values given in the following references:  
    $^a$ Masterson et al. 2005  \cite{Masterson05},  
    $^b$ Aharonian et al. 2005a \cite{Aharonian05a},
    $^c$ Aharonian et al. 2006a \cite{Aharonian06a},
    $^d$ Aharonian et al. 2006b \cite{Aharonian06b}. The X-ray luminosities
    - except for \psrkooka \ - are based on the power 
    law values given in the latest \inte \ Reference
    Catalogue Version 26 \cite{bird066}. ($^*$) Only a estimation for the
    X-ray flux and thus for the X-ray luminosity is possible due to the
    faint detection. The second VHE source in the Kookaburra region,
    G313.3+0.1, is not seen with \inte \ and therefore not listed in the table.}
  \label{tab:pulsarparameter}
  
  \begin{tabular}{p{85pt}||p{90pt}p{90pt}p{90pt}p{90pt}}\hline\noalign{\smallskip}\\
    \textbf{Name}                                          & PSR J0534+2200              & PSR J1513-5908              & PSR J0835-4510             & PSR J1420-6048             \\[3pt] 
                                                           & (Crab Nebula)               & (MSH 15-52)                 &                            & (G313.3+0.6)               \\[3pt] \tableheadseprule\noalign{\smallskip}
    \textbf{$P$ /[ms]}                                     & $33$                        & $151$                       & $89$                       & $68$                       \\[3pt] 
    \textbf{$\dot{P}$ / [s/s]}                             & $4.23 \cdot 10^{-13}$       & $1.54 \cdot 10^{-12}$       & $1.25 \cdot 10^{-13}$      & $8.32 \cdot 10^{-14}$      \\[3pt] 
    \textbf{$\mathrm{log}_{10}~\dot{E}$/[erg/s]}           & 38.7                        & 37.2                        & 36.8                       & 37                         \\[3pt] 
    \textbf{$\tau$ /[kyrs]}                                & 1.24                        & 1.55                        & 11.37                      & 13                         \\[3pt] 
    \textbf{distance / [kpc]}                              & 2.0                         & 4.4                         & 0.29                       & 7.69                       \\[3pt]
    \textbf{$\mathrm{log}_{10}~L_{VHE}$/[erg/s]}      & $34.4$ ($^a$)               & $34.6$ ($^b$)               & $32.9$ ($^c$)              & $34.9$ ($^d$)              \\[3pt] 
    \textbf{~~~(1TeV - 10TeV)}                             & $ $                         & $ $                         & $ $                        & $ $                        \\[3pt]
    \textbf{$\mathrm{log}_{10}~L_{X}$/[erg/s]}        & $36.6$                      & $35.2$                      & $32.7$                     & $\approx 34.6$ ($^*$)      \\[3pt]
    \textbf{~~~(20 - 40keV)}                               & $ $                         & $ $                         & $ $                        & $ $                        \\[3pt] \noalign{\smallskip}\hline
  \end{tabular}
\end{table*}

\subsection{PSR J1513-5908 (MSH 15-52)}
\label{subsec:2a}
\begin{figure}[t] \centering
\includegraphics[width=0.95\linewidth]{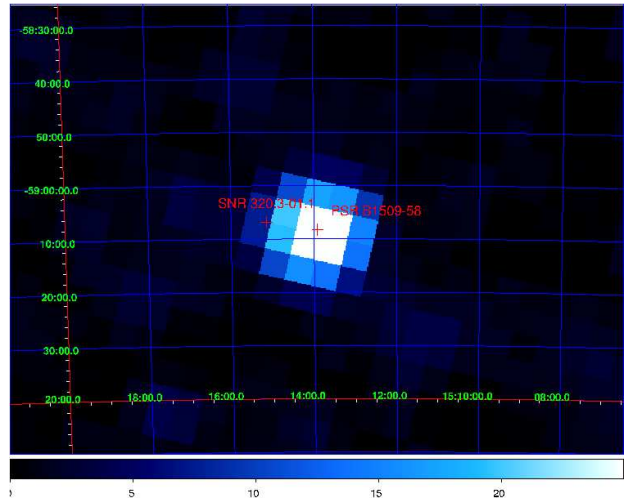}
\caption{\label{fig:mshinte} \psrmsh \ seen with \inte. The gray scale
indicates the significance in the range from $0-25~\sigma$. (See the Online
version for a color version of all figures.)} \end{figure}

\begin{figure}[t] \centering
\includegraphics[width=1.\linewidth]{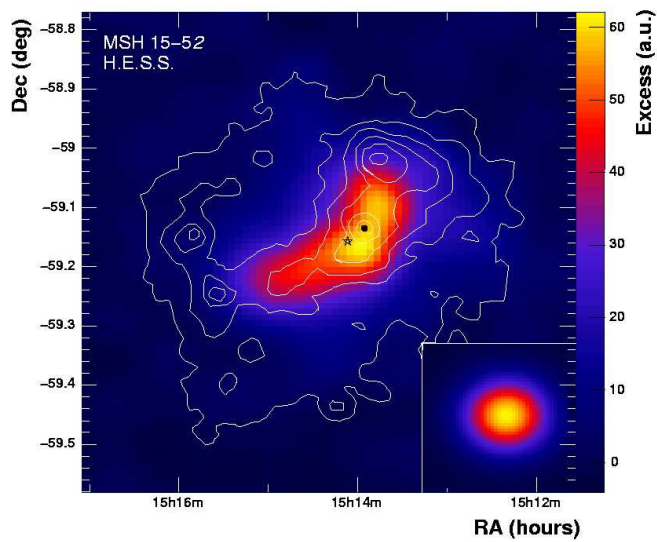}
\caption{\label{fig:hessj1514} \psrmsh \ seen with \hess \ (Figure from
\cite{Aharonian05a}). Overlaid (white contour) is the \rosat \ X-ray
($0.6-2.1$~keV) count rate.} \end{figure}

One of the young RPWNs with only an age of about 1500 years is the powerful
pulsar \ \psrmsh \ associated with the MSH 15-52 supernova remnant. The 150 ms
pulsar \psrmsh \ is seen in the X-ray, gamma-ray and radio energy band.  Fig.
\ref{fig:mshinte} shows the \inte \ mosaic for the region around \psrmsh.  For
comparison the VHE image for this region is shown in Fig.  \ref{fig:hessj1514}.
It is interesting to note, that the emission seen by \inte\ is possibly
spatially resolved with \inte\ to be extended (see also Terrier et al., these
proceedings).  

\subsection{PSR J0835-4510 (Vela)}
\label{subsec:2b}
\begin{figure}[t]
  \centering
  \includegraphics[width=0.95\linewidth]{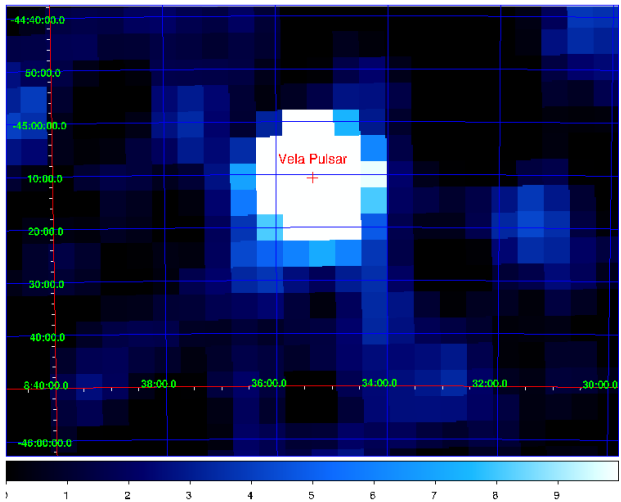}
  \caption{\label{fig:velainte} \vela \ seen with \inte. The significance mosaic for the
    energy band between $20-40$ keV has an exposure time of
    $\approx~1.3$~Msec. The scale is truncated below 0 and above
    10~$\sigma$.} 
\end{figure}

\begin{figure}[t]
  \centering
  \includegraphics[width=1.\linewidth]{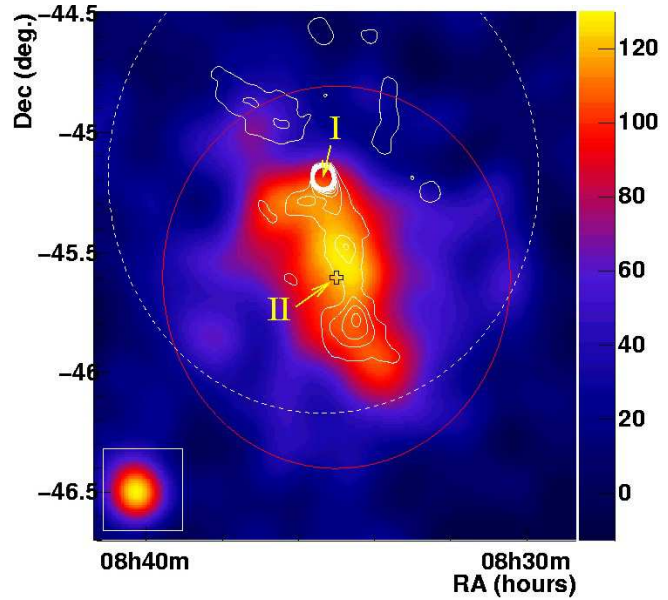}
  \caption{  \label{fig:velahess} \psrvela \ seen with \hess \ (Figure from
  \cite{Aharonian06a}.) \psrvela \ is located at position I. The white
  contours belong to the X-ray emission observed by \rosat.}
\end{figure}

The signal in the \inte \ \ibis \ image of the \vela is about
44.2 $\sigma$ (Fig. \ref{fig:velainte}). Most of the emission above
$20-60$~keV is unpulsed (Hermsen et al.,
priv. communication). Fig. \ref{fig:velahess} shows the \vela \ seen with
\hess \ \cite{Aharonian06a}. No indication for an extended emission in hard
X-rays from the Vela X region is apparent (see also Horns et al., these
proceedings). Combining spectral data of the ASCA satellite with the
non-detection, we conclude that there exists a spectral cut-off in the
energy range between $10-20$~keV.


\subsection{PSR J1420-6048 (Kookaburra)}
\label{subsec:2c}
\begin{figure}[t]
  \centering
  \includegraphics[width=1.\linewidth]{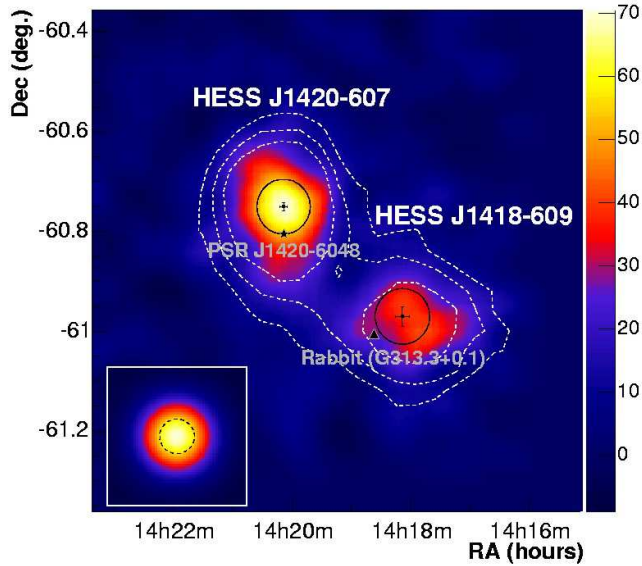}
  \caption{  \label{fig:kookahess} The \kooka \ region seen with \hess \ (Figure from
    \cite{Aharonian06b}). Shown is the smoothed excess map (gray scale or
    color in the Online Version) overlaid by the significance (contour).
    The inlaid box in the bottom left shows the point spread function for
    the \hess \ observation.}
\end{figure}

The discovery of the TeV emission, associated with the two radio wings of
the Kookaburra complex with \hess \ \cite{Aharonian06b}, confirms their
non-thermal nature and establishes their connection with the two X-ray
pulsar wind nebulae candidates (Fig. \ref{fig:kookahess}). As an
explanation for the Very High Energy gamma-rays inverse Compton scattering
of accelerated electrons on the Cosmic Microwave Background is assumed. The
\inte \ mosaic for the energy range of $20-40$ keV shows a faint signal of
$5 \sigma$ at the position of the \psrkooka. The pulsar  
\psrkooka \ belongs to the class of middle-aged pulsars where 
the RPWN interacts with the reverse shock of
supernova remnant. Interestingly, there is no \inte\ counterpart to the
``Rabbit'' RPWN candidate \cite{Roberts99,Ng05}. This object is presumably
powered by a young (as yet not clearly detected pulsar). While the two TeV
plerions show a similar
morphology and energy spectrum, their X-ray properties appears to be very
different. 

\section{Summary}
\label{summary}

Our preliminary study of TeV plerions with \inte \ indicates that some
of the pulars associated with these TeV plerions are \inte sources including
young and middle aged RPWN. The \inte \ observations are complementary to
the TeV observations as the unpulsed hard X-rays are presumably tracing the
on-going particle acceleration at the relativistic standing shock. In the
frame of ion induced acceleration, the observed hard X-ray spectrum
constrains the bulk Lorenz factor of the relativistic wind and its
composition.

\begin{acknowledgements}
We acknowledge the support of the Deutsches Zentrum f\"ur Luft- und
Raumfahrt under grant number 50OR0302. This work is based on observations
with \textsl{INTEGRAL}, an European Space Agency (ESA) project with
instruments and science data centre funded by ESA member states
(especially the PI countries: Denmark, France, Germany, Italy,
Switzerland, Spain), Czech Republic and Poland, and with the
participation of Russia and the USA.
\end{acknowledgements}

\end{document}